# Cellular dynamics of host - parasitoid interactions: insights from the encapsulation process in a partially resistant host

Samuel Gornard[1], Florence Mougel[1,*], Isabelle Germon[1], Véronique Borday-Birraux[1], Pascaline Venon[1], Salimata Drabo[1], Laure Kaiser[1]

[1] EGCE, Université Paris-Saclay, CNRS, IRD, UMR Évolution, Génomes, Comportement et Écologie, 91190 Gif-sur-Yvette, France

* Corresponding author. E-mail address : florence.mougel-imbert@universite-paris-saclay.fr (F. Mougel).

## Highlights

- Ovolarval development of parasitoid inside permissive host is established
- Encapsulation of parasitoid is delayed compared to that of inert beads
- Parasitoid capsules do not melanize compared to bead capsules
- Parasitism does not impact host melanization and encapsulation abilities
- Total hemocyte count remains unaffected by parasitism

## Article info

## Abstract

*Cotesia typhae* is an eastern African endoparasitoid braconid wasp that targets the larval stage of the lepidopteran stem borer, *Sesamia nonagrioides*, a maize crop pest in Europe. The French host population is partially resistant to the Makindu strain of the wasp, allowing its development in only 40% of the cases. Resistant larvae can encapsulate the parasitoid and survive the infection. This interaction provides a very interesting frame for investigating the impact of parasitism on host cellular resistance. We characterized the parasitoid ovolarval development in a permissive host and studied the encapsulation process in a resistant host by dissection and histological sectioning compared to that of inert chromatography beads. We measured the total hemocyte count in parasitized and bead-injected larvae over time to monitor the magnitude of the immune reaction. Our results show that parasitism of resistant hosts delayed encapsulation but did not affect immune abilities towards inert beads. Moreover, while bead injection increased total hemocyte count, it remained constant in resistant and permissive larvae. We conclude that while *Cotesia* spp virulence factors are known to impair the host immune system, our results suggest that passive evasion could also occur.

# Graphical abstract

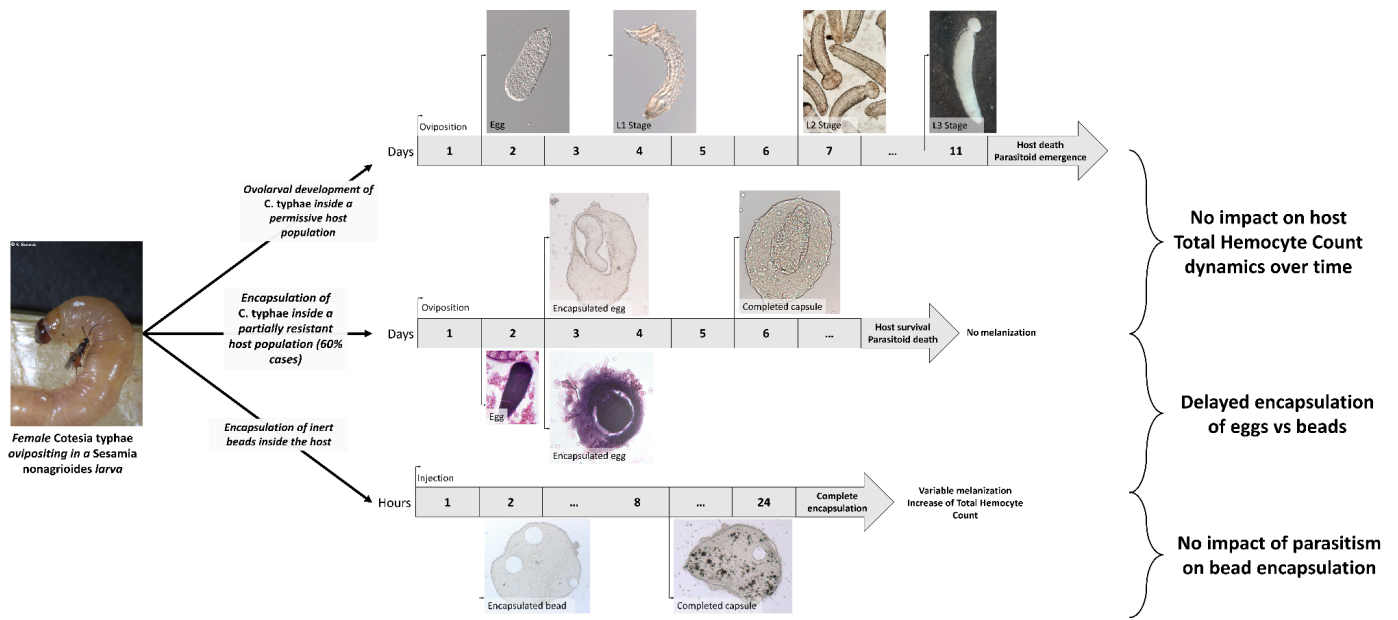

1. Introduction

Parasitoid insects can lead to important immune challenges in their arthropod hosts, as they need to lay their eggs on or in them in order to complete their life cycle (Eggleton & Belshaw, 1992). A broad majority of parasitoids belong to the Hymenoptera and Diptera order, the former commonly referred to as "parasitoid wasps" (Belshaw et al., 2003; Foottit & Adler, 2017). The parasitoid's offspring will develop by consuming the host's resources, which will, in most cases, lead to the host's death (Godfray, 2004), distinguishing parasitoids from regular parasites. Endoparasitoid wasps oviposit directly inside their host, and this constitutes an interesting case to study arthropods' immune responses, as the host's immune system must face the intrusion of foreign bodies (Godfray, 2007; Gullan & Cranston, 2014). This intrusion is especially important in the case of gregarious wasps, which can lay up to a hundred eggs inside a single host. The ability to negate the parasitoid's development is called host resistance, which resorts to an elaborate innate immunity (Ali Mohammadie Kojour et al., 2020), even if arthropods may show some kind of adaptive immunity (Cooper & Eleftherianos, 2017; Kurtz, 2005). This innate immune system acts by combining two ways of action, humoral and cellular. Humoral immunity is based on the production of antimicrobial molecules, either peptides or cytotoxic compounds, aimed at killing pathogens (Dunn, 1990; Manniello et al., 2021). Cellular immunity resorts to cells called hemocytes, that either circulate freely in the hemolymph or stick to the walls of tissue that might be in contact with pathogens (Hillyer & Strand, 2014; Jiravanichpaisal et al., 2006). While bacteria or fungi are usually phagocytosed or enclosed in nodules (Lackie, 1988), larger bodies are encapsulated (Lavine & Strand, 2002), and this constitutes the main defense against parasitoids. Detection of the eggs or larvae as non-self by the immune system usually triggers the recruitment, proliferation, and differentiation of hemocytes, which will form a multi-layer shell tissue around the invader and will often produce toxic compounds and melanin in addition to preventing interactions with other tissues (Carton et al., 2008; Nappi et al., 1995), like feeding. While eggs and larvae can both be targeted by encapsulation (Alleyne & Wiedenmann, 2001; Godfray, 2007), the egg stage is the most commonly encapsulated (Fellowes & Godfray, 2000), as the larvae can manage to evade encapsulation by moving or molting (Sullivan & Neuenschwander, 1988). Hence, the targeted encapsulation stage can modulate this resistance phenomenon. In insect hosts, cells involved in encapsulation have already been described in Diptera and Lepidoptera. In Diptera, such as mosquitoes or *Drosophila* flies, recognition of a foreign body (such as chromatography beads) as non-self will trigger the proliferation of circulating hemocytes by cell division, differentiation from pro-hemocytes, and production by dedicated hematopoietic organs (Carton & Nappi, 1997). In this order, encapsulation is mediated by lamellocytes (Hillyer & Strand, 2014; Lemaitre & Hoffmann, 2007), which are hemocytes specific to this taxon. In some Lepidoptera, such as *Chrysodeixis includens* (former *Pseudoplusia includens*, Lepidoptera: Noctuidae), granulocytes are the first cells to attach to the beads, followed by several layers of plasmatocytes and by a final layer of granulocytes, which role is to end the encapsulation process by mimicking an intact basal membrane (Carton et al., 2008; Strand, 2008). In others, such as *Manduca sexta*, the dynamic is less clear, with a first layer of plasmatocytes followed by random recruitment of both plasmatocytes and granulocytes (Carton et al., 2008; Wiegand et al., 2000). After successful encapsulation, parasitoid death can occur either by asphyxia or poisoning by cytotoxic compounds produced by the cells building the capsule (Blumberg, 1997; Lavine & Strand, 2002). In most cases, the hemocytes will also release melanin, blackening and hardening the capsule, but that is not systematic (Lackie, 1988; Pech & Strand, 1996).

To escape encapsulation, parasitoid wasps can either evade the immune system by preventing detection of the eggs (Huw Davies & Vinson, 1986; Prevost et al., 2005) or disrupt it to allow development. The latter strategy is performed by injection of various factors alongside the eggs: venom (Mrinalini & Werren, 2016; Yu et al., 2023), polyDNAviruses (Dupuy et al., 2006; Strand & Burke, 2014, 2020), ovarian fluid (Mabiala-Moundoungou et al., 2010; Salvia et al., 2022, 2023) or Virus-Like Particles (VLP, Feddersen et al., 1986; Morales et al., 2005). Some of them are produced during the embryonic development of the offspring, which is the case for the teratocytes (Gao et al., 2016; Gu et al., 2019; Salvia et al., 2019). The parasitoid's ability to thwart its host's immune system is called virulence. Such immunosuppression can be linked to drops in total hemocyte count, disruption of cell spreading and signaling cascades, or inhibition of the phenoloxidase, an enzyme responsible for the production of cytotoxic compounds and melanin (Asgari, 2006; Colinet et al., 2009; Ibrahim & Kim, 2006). Variations of virulence on the same host between parasitoid strains or close species have already been observed. This is the case for the couple between sister species *Leptopilina heterotoma* and *L. boulardi* (Hymenoptera: Figitidae), parasitizing *Drosophila melanogaster* (Diptera: Drosophilidae) (Cavigliasso et al., 2019; Colinet et al., 2013), or between *Cotesia sesamiae* (Hymenoptera:

Braconidae) strains *inland* or *coast*, and the host *Busseola fusca* (Gitau et al., 2007; Mochiah et al., 2001). Such cases are great opportunities to study genes and proteins involved in parasitoid virulence and host resistance, as a molecular comparison between the couples can draw candidates that can be characterized afterward. However, as the dynamics of host resistance are variable among species, characterizing the precise kinetics of this interaction is essential prior to any fine molecular analysis of genetic basis.

Another species where differences in virulence between strains have been reported is *Cotesia typhae* (Hymenoptera: Braconidae, Fernández-Triana), a gregarious larval parasitoid of lepidopteran stem borers specialized on *Sesamia nonagrioides* (Lepidoptera: Noctuidae) (Kaiser et al., 2017). *C. typhae* is only found in East Africa, whereas its host has become a major crop pest in southern Europe (Brookes, 2009; Meissle et al., 2009). Thus, *C. typhae* could potentially be used as a biological control agent there. Two strains of *C. typhae* are currently reared in the laboratory and are named after the Kenyan location they were initially collected from, Kobodo and Makindu. Both display an important parasitism success when egg-laying in larvae from the Kenyan population of *S. nonagrioides*, with 94% of host larvae successfully parasitized for Makindu and 88% for Kobodo. However, when parasitizing the French population, this success significantly drops for the Makindu strain, reaching 30%, but stays at the same level for the Kobodo strain (94%) (Benoist et al., 2017, 2020). This difference in the interaction between the host and the parasitoid can be explained by a lower virulence in Makindu parasitoid combined with an increased resistance of the French host, which would lead to a successful immune reaction of the host and potential encapsulation of the parasitoid's eggs or larvae.

The case of *C. typhae* and *S. nonagrioides* is particularly interesting as the European host population diverged from the African one 180 000 years ago (Kergoat et al., 2015), which resulted in a genetic differentiation between the two populations (Moyal et al., 2011). The European population might have evolved in response to local parasitoids, predators, or pathogens, in adaptation to temperate climate, or even have undergone a bottleneck effect. *C. typhae* has not followed its host in Europe and has thus recently been in secondary contact with the French host, for scientific purposes (Benoist et al., 2017; Kaiser et al., 2023). Unexpectedly, while the Kobodo strain is still highly virulent on the French host, the Makindu strain displays a parasitism success that is not constant. In this system, we witness intraspecific and intrapopulation differences in both host resistance and parasitoid virulence, contrary to other species couples cited above. Study of the particular combination where resistance occurs will then enlighten the comprehension of cellular immunity processes and can give insights into how host-parasitoid interactions can switch from permissive to forbidding.

To better understand the ongoing interaction, it is essential first to characterize the situations leading to encapsulation. In the specific case of true host resistance (French *S. nonagrioides* parasitized by Makindu *C. typhae*), it is then necessary to describe the cellular mechanisms and their temporal dynamics. Total hemocyte count (THC), by reflecting the number of circulating hemocytes within the hemolymph, can also cast a light on ongoing immune reactions.

In this study, we have checked the presence or absence of capsules following parasitism in the four possible host-parasitoid interactions. Then, we have characterized the dynamic of the encapsulation process in resistant French *S. nonagrioides* larvae and compared it to the encapsulation of chromatography beads, while measuring total hemocyte count within the hemolymph.

## 2. Material & Methods
### 2.1. Host and parasitoid rearing

*Sesamia nonagrioides* larvae are reared in the lab on an artificial soy-corn-yeast diet (adapted from Overholt et al., 1994). Nine-day-old eggs displaying a characteristic gray color are sterilized with Cooper© Dakin's solution and placed at the surface of the medium. The larvae then develop for 5 to 6 weeks at 26°C, 70% relative humidity, and a 16/8 light/dark cycle before the chrysalids are collected and placed in a cage (21°C, 70% RH, 16/8 light cycle) where adults can reproduce and lay eggs on paper stems. A French and an African strain are reared. The French one (SNF) originates from several localities in southwest France where *S. nonagrioides* causes yearly damage to maize. The African one (SNK) originates from two Kenyan localities in south-east Kenya, Makindu and Kabaa.

*Cotesia typhae* adults are kept in plastic boxes with access to water and sugar supply (honey drops) at 70% RH and 12/12 light/dark cycle. One to three-day-old females are used for egg-laying. Three-week-old SNK larvae (5th instar: L5) are used for parasitoid reproduction, as both lineages display a high success in this host population. Larvae to be parasitized are beforehand fed on fresh maize stems for two days before parasitism. Each larva is presented to one female wasp, and successful oviposition is attested by the experimenter. Parasitized larvae are then placed back on an artificial medium for 12 days under classical rearing conditions. They are then removed and placed on tissue paper so that *C. typhae* larvae can burst out of the host on a dry substrate, where they can spin their cocoon safely. This occurs synchronously, resulting in a mass of cocoons surrounding the host body. Cocoon masses are then collected and placed in plastic boxes for adults. Adult emergence is monitored daily so that adults spend one day at 27°C for mating and one at 24°C before being placed at 21°C for maximum survival until parasitism.

### 2.2. Parasitoid development

Ovolarval development of the Makindu strain inside a permissive host was assessed by dissecting Kenyan *S. nonagrioides* (SNK) larvae parasitized by the Makindu *C. typhae* strain, under the same conditions as for the rearing. Five individuals were dissected per time step, at 0 to 11 days post egg-laying, for a total of 55 dissected larvae. Euthanasia was performed by placing the insects for 30 minutes in a 15 mL falcon tube containing a diethyl-ether-soaked cotton ball. This method was considered harm-free and nondestructive for the tissues, contrary to a cold shock. Dead larvae were surface-sterilized with 96° ethanol, placed on a glass petri dish, and had their head and tail removed. They were then cut just below the surface of the cuticle, alongside an anteroposterior line located on the dorsal side of the animal, just above the spiracle line. The animal's inner cavity was washed with sterile phosphate buffer solution (PBS) 1X, and *C. typhae* individuals found on the bottom of the petri dish were collected with a micropipette under a Leica MZ7.5 Stereomicroscope and stored in PBS 1X for further examination.

### 2.3. Encapsulation abilities assessment

Encapsulation abilities of *S. nonagrioides* populations were first determined by dissecting fifth instar (L5) larvae 4 days after parasitization. Fifty larvae of each population were parasitized by the Makindu parasitoid strain, for a total of 100 individuals. Due to the higher virulence hypothesized in the Kobodo strain, 92 and 125 larvae of SNF and SNK host populations were parasitized. Euthanasia and dissections were performed as described before, and retrieved parasitoid larvae were classified as unencapsulated, partially encapsulated (mix of encapsulated and free larvae) or totally encapsulated. Larvae that were shown to be devoid of any parasitoid after close inspection were also reported.

### 2.4. Dynamic of encapsulation

L5 SNF larvae were parasitized each by one Makindu wasp. In order to get capsules at different stages of formation, several parasitized SNF were dissected until the obtention of four resistant individuals (showing capsules) per time step, at 2, 3, 4, and 5 days post egg-laying. In addition, eight parasitized SNF larvae were dissected one day post egg-laying to ensure the absence of encapsulation at this time step. Euthanasia, dissection, and recuperation of capsules were performed as described before. The content of the capsules was then observed under light microscopy between slide and cover slip. Three characteristics were recorded: the presence of hemocytes (capsule formation), the presence of final basal membrane (capsule termination), and melanization (present or not).

Encapsulation and melanization processes in *S. nonagrioides* free of parasitoid virulence factors were observed by injecting L5 SNF larvae with foreign bodies able to trigger an immune reaction. We used Sephadex G-50 beads (SIGMA, Havard et al., 2014) suspended in sterile PBS 1X. Larvae were surface-sterilized with Dakin's and placed under aseptic conditions, within the sterility cone of a Bunsen burner. Between 60 and 80 beads (20 µL PBS-bead solution) were injected with an insulin needle (MYJECTOR Terumo 1 mL 0.33x12 mm) in the larvae's hemolymph.

Four Sephadex-injected larvae were dissected per time step, at 1, 2, 3, 4, and 8 hours, as well as 1 to 4 days post-injection, for a total of 36 injected larvae. Hemocyte-covered beads were retrieved in the petri dish and observed under light microscopy. Capsule characteristics were recorded as before.

The absence of mortality induced by the injection, either by bacterial infection or mechanical stress, was assessed by injecting 25 L5 larvae with 20 µL of sterile PBS 1X free from beads.

Finally, impact of parasitism on bead encapsulation was assessed by injecting Sephadex beads in L5 SNF larvae parasitized by Makindu wasps. Larvae were dissected at key time steps of bead and egg encapsulation, 4, 48 and 96 hours after parasitism plus injection. An equivalent amount of larvae solely injected with beads were dissected as control. Fifteen larvae per time step and condition were dissected. The beads were retrieved as described before, and the same characteristics were recorded. Larvae parasitism was attested by the presence of the parasitoid (egg or larva) and their encapsulation or melanization levels were also recorded.

### 2.5. Histological approach

Makindu-parasitized SNF larvae were sampled at 24, 36, or 48 hours post egg-laying (3 individuals per step of time). One control individual was sampled immediately after oviposition. Euthanasia was performed as described above. Because of the host size and the impermeability of its cuticle, most incubation times and volumes were greatly increased compared to a classical embedding protocol. Dead larvae were fixated in 50 mL of PBS containing 4% paraformaldehyde (PFA) for 1 hour at room temperature (RT) under agitation, and then for three days at 4°C. After washing in PBS 1X, samples were progressively dehydrated at RT in graded ethanol series under agitation and stored overnight in absolute ethanol at 4°C. They were transferred in butanol for seven days at RT and then embedded in paraplast for 12 hours at 58°C. Paraplast was replaced four times, with the last bath taking place under vacuum. Transversal sections of 17µm thick were made with Leica RM 2255 Rotary Microtome. Histological sections were counterstained with Mayer's hematoxylin solution (1 g/l, SIGMA) and eosin Y solution 0.5% aqueous (SIGMA), then mounted with Eukitt® mounting media. Before embedding, each sample was cut for 1,5 to 2,5 cm around the stinging point, representing 880 to 1500 sections per sample.

### 2.6. Total hemocyte count

To control for the physiological effect of our treatments, we measured the total hemocyte count (THC), expressed as the number of hemocytes (per milliliter of hemolymph) of SNF larvae, 2 and 24 hours after injection of Sephadex beads or injection of sterile PBS. The effect of parasitism on the THC was measured 2, 24, and 96 hours after parasitization by Makindu *C. typhae*. For the latter condition, both resistant and permissive SNF individuals were sampled. After hemolymph sampling, larvae were dissected to determine whether they were resistant (presence of capsules) or permissive (presence of developed parasitoid larvae). Twenty microliters of hemolymph were sampled from cold-anesthetized larvae as described by Ibrahim and Kim in 2006. Collected hemolymph was placed in a microtube with 100 µL of AntiCoagulant Buffer (AcB, Mead et al., 1986) before freezing at -20°C. Measures of THC were done with a Malassez cell counting slide (Brand™). Eight individuals per condition and time step were assessed. Counts were also carried out on untreated control L5 larvae.

Because age can influence resistance capabilities via hemocyte density, with older larvae (that is, prior to entry into the pre-nymph stage) tending to have more circulating hemocytes (Gardiner & Strand, 2000), we used caterpillars that were all the same age (± 3 days) and stage (L5) to minimize this bias. Moreover, as the growth rate, and consequently the THC density, may vary among individuals according to rearing population density and access to nutrient medium (Mohamed & Amro, 2022; Vogelweith et al., 2016), caterpillars with similar size were chosen for the experiments.

### *2.7. Statistical analysis*

The results of parasitism and/or bead injection revealed by dissection were compared using the Chi-square test or Fisher's exact test when the numbers were too low. This was followed, if needed, by a post-hoc test (Ebbert, 2022).

THC results were compared using a one-way ANOVA followed by a Tukey HSD post-hoc test.

Analyses were performed with R software, version 4.1.2 (R Core Team, 2018).

# 3. Results

## 3.1. Ovolarval development of Cotesia typhae

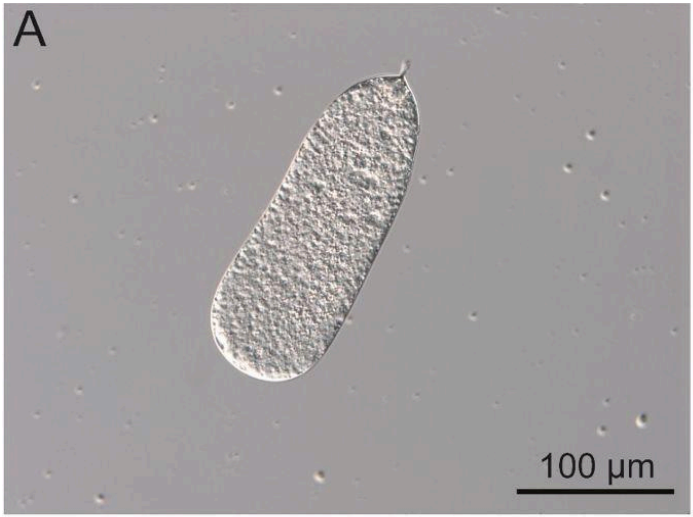
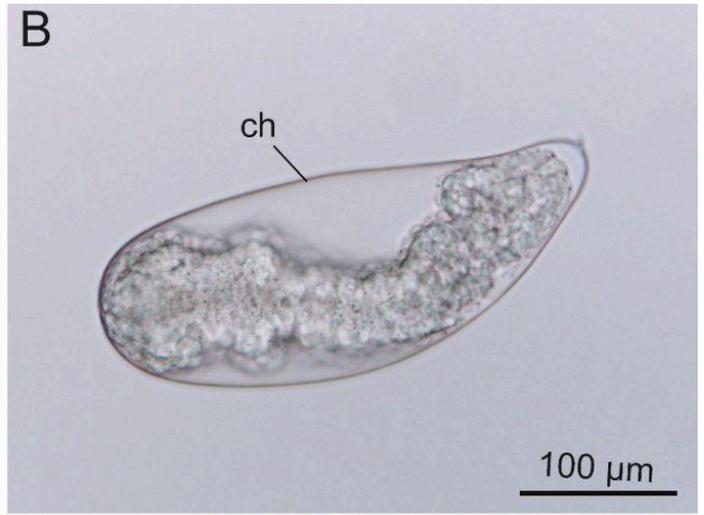
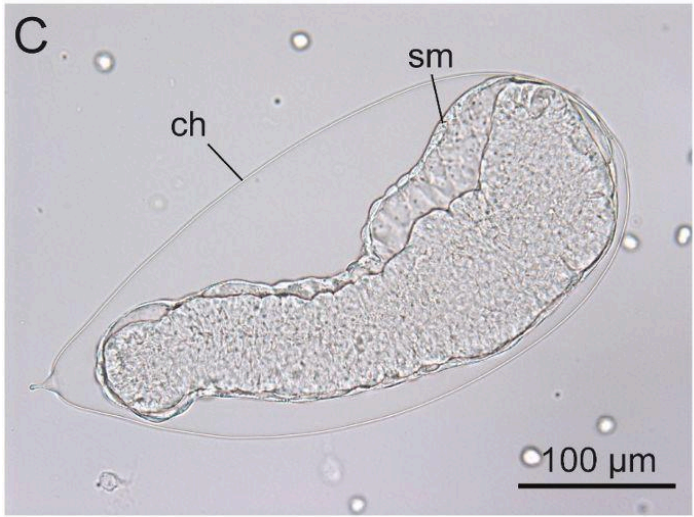
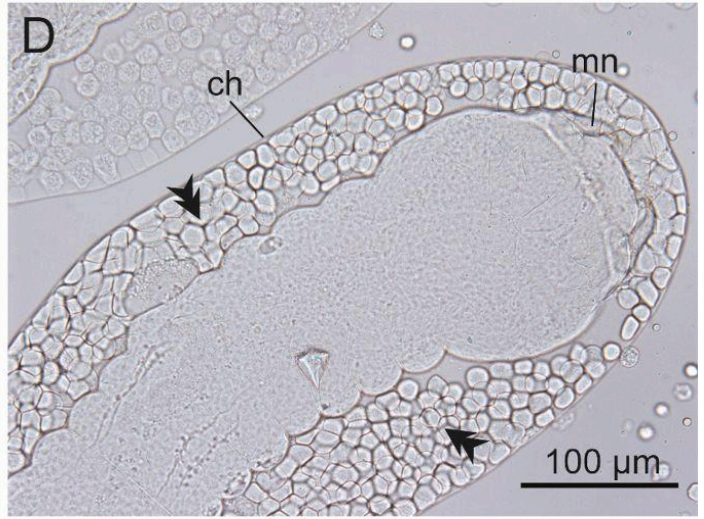
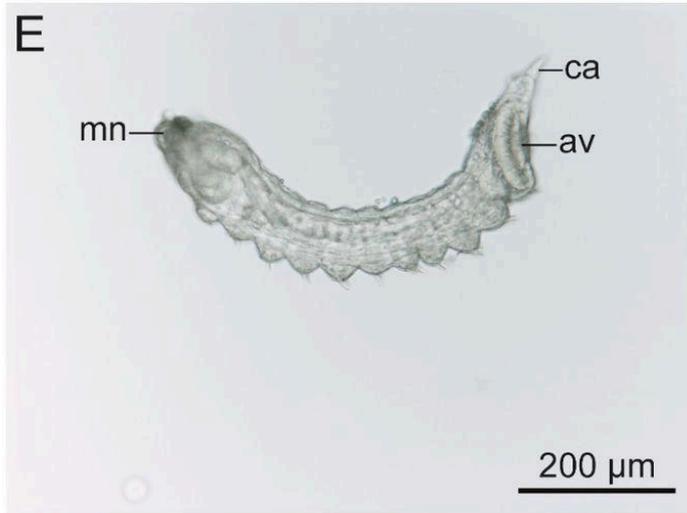
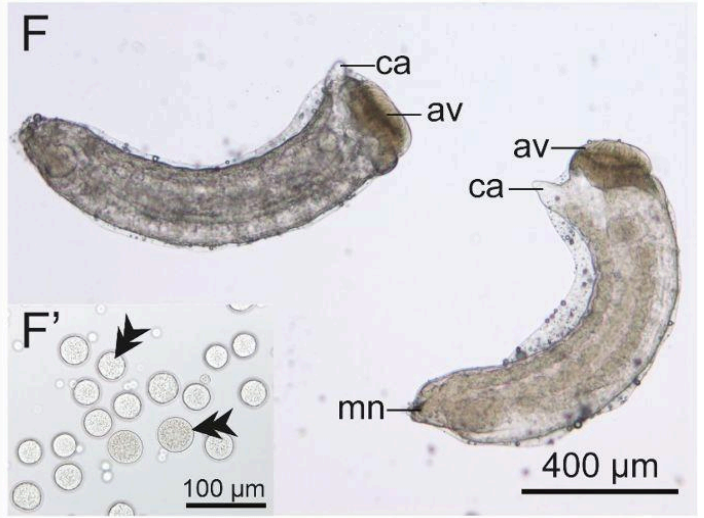
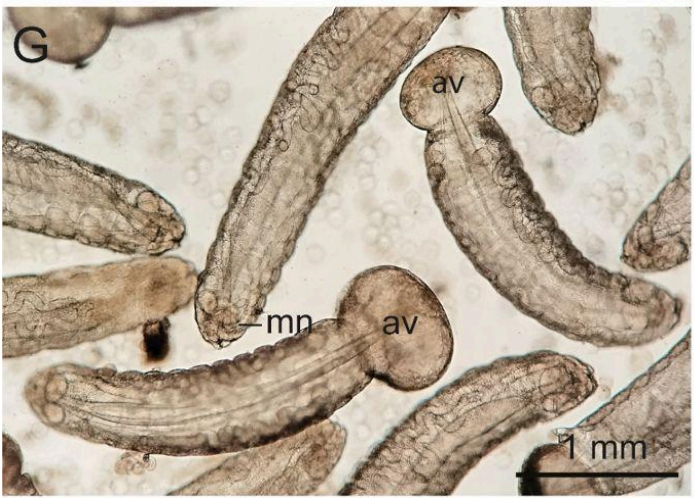
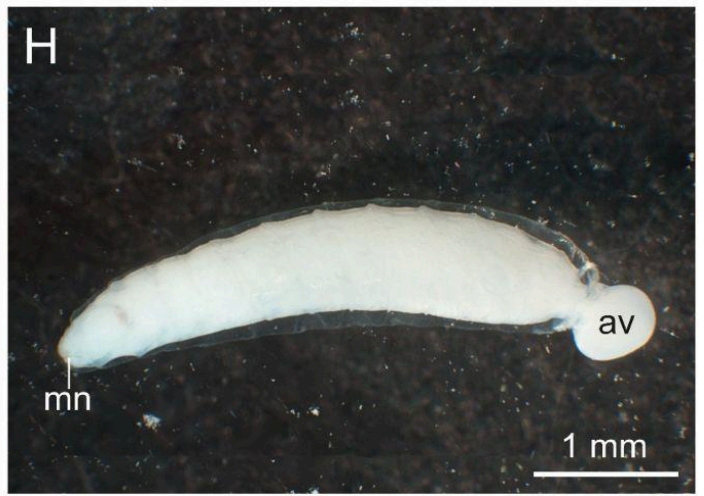

Figure 1: Ovolarval development of Makindu *C. typhae* strain in the permissive Kenyan host. Nomarski light photomicrographs of (A) Embryo stage, one day after oviposition (AO). Light photomicrographs of (B) Pre-hatched embryo, two days AO, inside the egg chorion (ch). (C) Pre-hatched embryo, two days AO, with delaminating serosal membrane (sm). (D) Late-eclosing embryo, three days AO, with chorion full of teratocytes (double arrowheads) and visible mandibles (mn). (E) First instar larva, three days AO, with anal vesicle (av) developing at the basis of the caudal appendage (ca). (F) First instar larva, five days AO. (F') Teratocytes retrieved in the host's hemolymph, five days AO. (G) Second instar larvae, six days AO. (H) Third instar larvae, eleven days AO. Pictures A to G were taken on a Leica DM6000 microscope and picture H on a Leica Z16 APO macroscope using a Leica DF490 and a Leica K5C color cameras respectively.

The different steps of ovolarval development of Makindu *C. typhae* strain have been observed in the Kenyan host at 27°C and are presented in Figure 1. The eggs of *C. typhae* are hydropic and have to absorb nutrients and water from their host's hemolymph to allow embryogenesis (Quicke, 2014). Their length (anteroposterior axis) is multiplied by 1.8 during embryogenesis (Fig. 1, A, B, C), similar to the development of *C. vestalis* eggs (Yu et al., 2008). After two to three days of embryogenesis (Fig. 1, A, B), eclosion takes place. First-instar larvae (Fig. 1, E, F), characterized by a caudal appendage (ca), thrive for about six days during which the anal vesicle (av) starts developing. This everted rectum is located at the posterior end of the larvae, next to the caudal appendage. It is even more developed in the second instar (Fig. 1, G), which is deprived of caudal appendage. The anal vesicle likely plays a role in respiration, nutrition, and excretion (Quicke, 2014). Third-instar larvae (Fig. 1, H) are not transparent anymore, and their anal vesicles are beginning to regress, similar to *C. flavipes* larval development (Pinheiro et al., 2010). Mandibles (mn) were clearly present in every instar, indicating their possible use for eclosion, feeding, and egression. Egression takes place between twelve and thirteen days post-oviposition.

Observation of embryos soon before hatching showed that teratocyte formation occurred in *C. typhae* (Fig. 1, C, D), as in other Braconidae (Quicke, 2014). Consistent with what Tremblay and Calvert (1971) reported in the Microgastrinae group, *C. typhae* teratocytes seem to derive from a serosal embryonic membrane that separates from the embryo through delamination (Fig. 1, C). Late-eclosing individuals could be observed inside the egg chorion, with grown teratocytes ready to burst out after eclosion (Fig. 1, D). Dissection of successfully parasitized *Sesamia nonagrioides* after more than three days AO revealed the presence in the hemolymph of giant, round, numerous cells that we hypothesize to be teratocytes having burst out of the chorion after eclosion (Fig. 1, F').

### 3.2. Encapsulation abilities assessment

| *Host* | SNK | | SNF | |
|---|---|---|---|---|
| *Parasitoid strain* | Kob | Mak | Kob | Mak |
| **Larvae** | 96.0 | 92.0 | 91.3 | 24.0 |
| **Capsules** | 0.0 | 0.0 | 0.0 | 54.0 |
| **Empty** | 4.0 | 8.0 | 8.7 | 22.0 |
| ***n*** | 125 | 50 | 92 | 50 |

Table 1: Percentage of L5 *S. nonagrioides* containing either *C. typhae* larvae, capsules, or that were devoid of any parasitoid, depending on host population (SNK and SNF) and parasitoid strain (Kobodo, Kob; Makindu, Mak). Sample sizes are reported in the bottom line.

The encapsulation abilities observed by dissection of SNK and SNF larvae parasitized by each parasitoid strain are reported in Table 1. Consistent with expectations, encapsulation was only observed in SNF hosts parasitized by Makindu (Pearson's exact test, p-value < 0.001). Among the 27 hosts that encapsulated Makindu, 16 succeeded with a complete encapsulation of all parasitoid larvae, 6 encapsulated more than half of them, and 5 less than half. No capsule was melanized at this time step. Those results suggest that SNK population is fully permissive to both *C. typhae* strains, while SNF population is permissive to Kobodo strain and partially resistant to Makindu strain.

Dissection also revealed that several host larvae in which oviposition clearly took place were devoid of parasitoids (Table 1, Empty). This phenotype was found in both strains, but occurred significantly more often in Makindu parasitizing SNF (Chi-square post-hoc test, p-value = 0.009). This result suggests it could be influenced by genetic background of the strain, host population or environmental factors, but also by specific host/parasitoid interactions. This also highlights that parasitized larvae that do not sire wasp offspring are not necessarily resistant, and that dissection is needed to attest host resistance.

Following this observation, successful oviposition was assessed by dissection for every tested larva in all the following experiments.

### 3.3. Dynamic of encapsulation

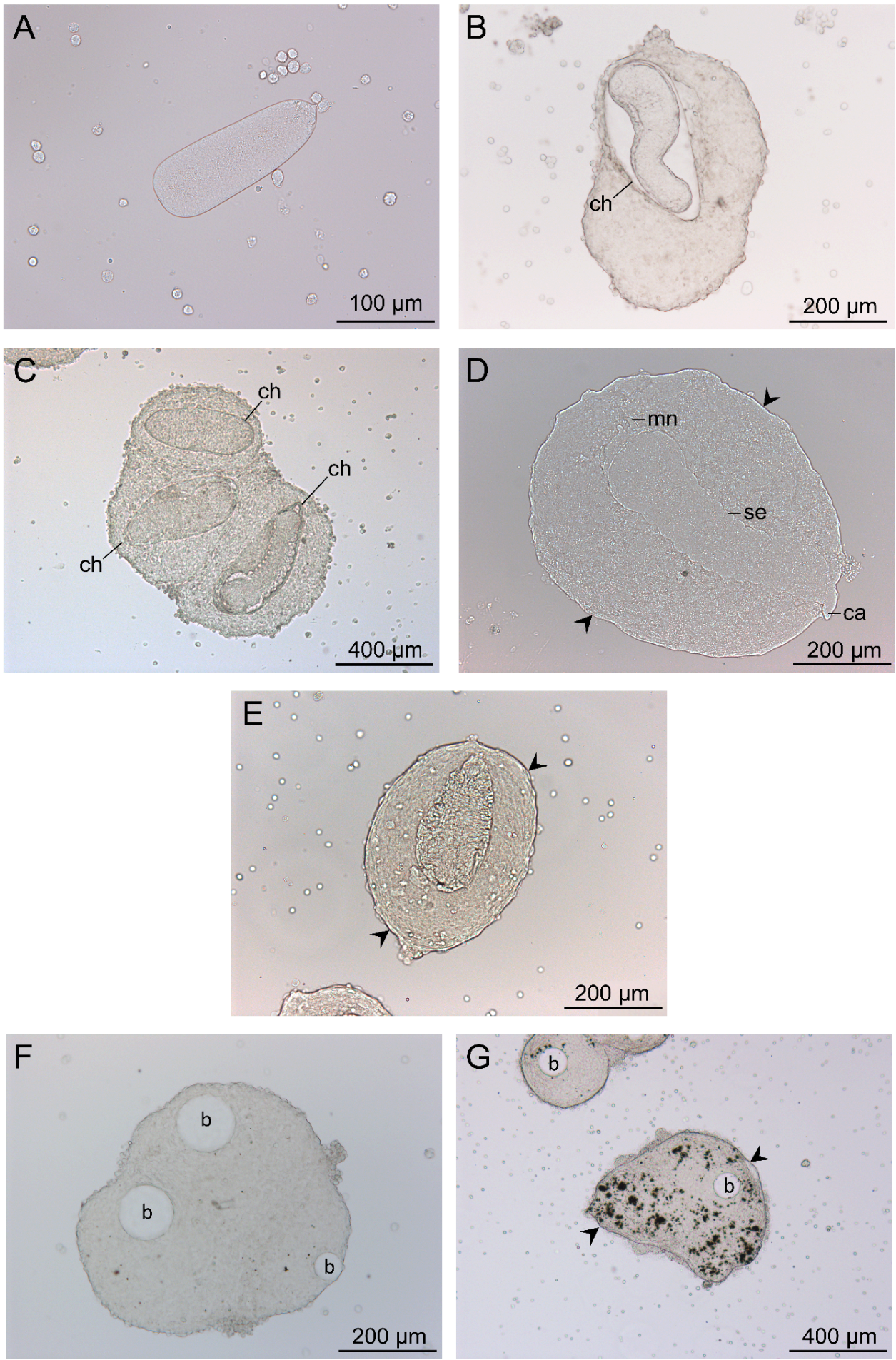

Fig 2: Encapsulation dynamics of *C. typhae* and Sephadex beads by *S. nonagrioides*. Light photomicrographs of: (A) Hemocyte-free egg, one day after oviposition (AO). (B) Unhatched embryo inside the encapsulated egg chorion (ch), two days AO. (C) Encapsulated larvae, three days

AO. (D) Completely encapsulated larva, four days AO, with visible basal membrane (arrowheads). Larval anatomy can be observed, with mandibles at the anterior part (mn), segments (se), and caudal appendage (ca). (E) Completely encapsulated larva, five days AO. (F) Encapsulated beads (b), one hour after injection, with terminal basal membrane under construction. (G) Completely encapsulated beads, eight hours after injection, with terminal basal membrane (arrowheads). The black dots correspond to the melanization of the capsule. Pictures were taken using a Leica DM6000 microscope using a Leica DF490 color camera.

Dissections of parasitized L5 French *S. nonagrioides* (SNF) larvae at various times after oviposition were conducted in order to establish the encapsulation dynamic of Makindu *C. typhae* offspring within resistant individuals. This dynamic was compared to that of inert chromatography beads, which act as a positive control for encapsulation. The different steps of encapsulation processes are presented in Figure 2.

The parasitoid's developmental stage targeted by encapsulation appeared to be the egg. One day after oviposition (Fig. 2, A), no hemocyte can be seen attached to the eggs, whereas at two days (Fig. 2, B), capsules around the eggs already present several cellular layers. Hemocytes start binding *C. typhae* eggs between one and two days AO and build the capsule for three days. Ovolarval development can be observed inside the egg chorion (Fig. 2, B to E). Four days AO, the capsules are terminated, as attested by the basal membrane meant to end the recruitment of new hemocytes (Fig. 2, D). Eclosion from the chorion takes place inside the capsule, and the larvae die, unable to pursue their development. Five days AO, the outer anatomy of the larvae trapped in the capsule is no longer distinguishable, as they are likely beginning to be digested by the capsule (Fig. 2, E).

Inert Sephadex beads were encapsulated more quickly than *C. typhae* individuals, as capsule formation is ongoing one hour after injection (Fig. 2, F) and can be terminated by the basal membrane after only 24 hours (Fig. 2, G). Melanisation can be observed on bead capsules (Fig. 2, G, black dots).

|  | Time step | 1h | 2h | 3h | 4h | 8h | 1d | 2d | 3d | 4d | 5d |
|---|---|---|---|---|---|---|---|---|---|---|---|
| **Parasitism** | Encapsulation | - | - | - | - | - | 0* | 4 | 4 | 4 | 4 |
|  | Basal membrane | - | - | - | - | - | 0* | 0 | 1 | 4 | 4 |
|  | Melanization | - | - | - | - | - | 0* | 0 | 0 | 0 | 0 |
| **Beads** | Encapsulation | 4 | 4 | 4 | 4 | 4 | 4 | 4 | 4 | 4 | - |
|  | Basal membrane | 1 | 3 | 1 | 2 | 3 | 2 | 4 | 4 | 4 | - |
|  | Melanization | 3 | 4 | 4 | 4 | 4 | 4 | 3 | 4 | 3 | - |

Table 2: Capsule characteristics retrieved by dissection of L5 French *S. nonagrioides* (SNF). For each time step and condition (parasitism by Makindu *C. typhae* strain or injection of 20 µL of beads-PBS solution), the number of host larvae out of four (or eight*) showing the mentioned characteristics is indicated. No larvae died in any condition.

The occurrence of different characteristics of capsule formation over four (or eight) caterpillars and at different time steps is summarized in Table 2. Consistently with the proportion of larvae expected to be permissive, around 40% of larvae dissected at least 48 hours post-parasitism had to be discarded as they did not display encapsulation. Those permissive hosts were not reported in Table 2.

Encapsulation of *C. typhae* offspring never occurred within less than two days after oviposition and was completed for every individual after four days. Melanization was never present in any of the capsules found in resistant SNF. Encapsulation of beads occurred in less than 1 hour after injection and was completed for every individual after 48 hours. Melanization was seen in almost every individual and appeared less than one hour after injection. However, it was never fully complete and only consisted of black dots sprinkled over the capsule (see Fig. 2, G). The number of dots ranged from 1-5 (capsules not considered "melanized," Fig. 2, F) to around a hundred. It varied between and within individuals and was not correlated to the age of the capsule.

|  | Time step | 4h | 48h | 96h |
|---|---|---|---|---|
| **Parasitism + beads** | Encapsulation | 15 | 15 | 15 |
|  | Basal membrane | 4 | 4 | 12 |
|  | Melanization | 5 | 7 | 12 |
| **Beads** | Encapsulation | 15 | 15 | 15 |
|  | Basal membrane | 6 | 7 | 9 |
|  | Melanization | 10 | 7 | 14 |

Table 3: Bead capsule characteristics retrieved by dissection of L5 French *S. nonagrioides* (SNF). For each time step and condition (parasitism by Makindu strain and injection of 20 µL of beads-PBS solution or injection alone), the number of host larvae out of 15 showing the mentioned characteristics is indicated. No larvae died in any condition. Numbers of SNF displaying fully encapsulated or melanized beads were compared to control at each time step, using a Chi-square or a Fisher exact test.

Characteristics of bead capsules retrieved in SNF larvae parasitized by Makindu are summarized in Table 3, and compared to bead encapsulation in control larvae. First, control larvae confirmed the dynamic of *S. nonagrioides* immune reaction, with bead encapsulation occurring in every individual as soon as 4 hours post-injection, and basal membrane and melanization present in most individuals 96 hours after injection.

Parasitism did not impact bead encapsulation, as all parasitized individuals could encapsulate beads at every time step, and did not impact the proportion of individuals with fully encapsulated beads (Chi-Square, p-values between 0.449 and 0.699). However, six individuals displayed, 4 hours after oviposition and injection, a small number of beads that were completely free of hemocyte, while all others were partially or completely encapsulated. Such beads were found in one individual at 48 and 96 hours post-oviposition plus injection, and were never found in control individuals. Parasitism did not impact bead melanization levels (Chi-Square, p-values between 0.144 and 1)

Bead injection seemed to have an impact on parasitoid encapsulation, as no individual dissected 48 hours after oviposition plus injection displayed encapsulated larvae, unlike what is reported in Table 2. Nine out of 15 individuals dissected 96 hours after oviposition plus injection displayed encapsulated parasitoid eggs and larvae, but unlike what is reported in the phenotyping experiment and in Table 2, 8 of them were melanized.

### 3.4. Histological investigation of the encapsulation process

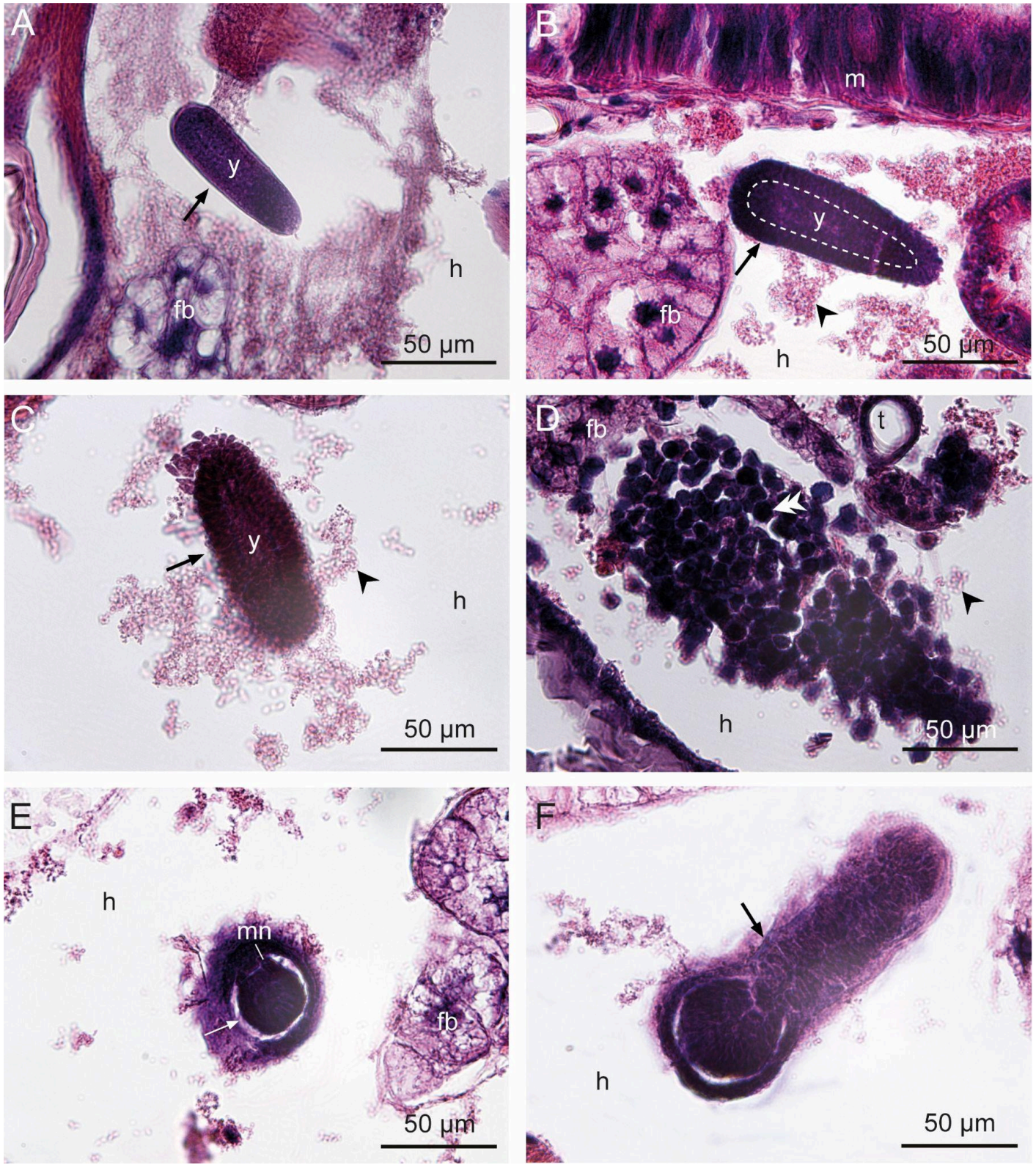

Fig 3: Encapsulation dynamic of *C. typhae* eggs inside *S. nonagrioides* larva. Light photomicrographs of eosin-hematoxylin-colored histological cross sections of the host. Arrows point out parasitoids.
(A) Lateral section of an egg, immediately after oviposition (AO). (B) Lateral section of a cellular blastoderm embryo, 24h AO, surrounded by host hemocytes aggregates (arrowhead). Yolk is visible in the middle (surrounded by dotted line), with a single-layer cell jacket in the periphery. (C) Lateral view of an embryo, 36h AO, surrounded by host hemocytes. (D) Lateral view of a torn egg chorion, 36h AO, with exposed teratocytes (double arrowhead). (E), (F) Encapsulated *C. typhae* larvae, 48h AO. (E) Cross section of the head with visible mandibles (mn). (F) Frontal section. Pictures were taken using a Leica DM6000 microscope using a Leica DF490 color camera.
 fb: fat body, h: hemocoel, m: mesenteron, t: tracheole, y: yolk

Histological cross-sections on L5 *S. nonagrioides* larvae parasitized by Makindu *C. typhae* were performed to characterize better the encapsulation process related to the wasp's development. Eggs detected through this method were observed to be free roaming in the hemocoel of the host and not particularly attached to or concealed in any specific host tissue. They were spotted in all individuals sampled at 0 and 24 hours post-oviposition, two sampled at 36 hours, and one sampled at 48 hours. Another individual sampled at 48 hours exhibited encapsulated eggs. Two individuals were devoid of *C. typhae* due to the absence of egg-laying during oviposition.

Early embryogenesis of the wasp can be observed through the sections. Immediately after oviposition (AO) (Fig. 3, A), only yolk can be seen inside the chorion. The oviposition triggers embryogenesis, and at 24 hours AO, embryos have reached the cellular blastoderm stage of their development (Fig. 3, B). At 36 hours AO, gastrulation is ongoing (Fig. 3, C), and teratocytes are visible (Fig. 3, D). This result widens the time window when the teratocytes are produced, as they can be formed between 24 and 48 hours AO.

Aggregation of hemocytes is noticeable on sections at 24 and 36 hours AO (Fig. 3, B, C), which is not the case immediately AO. This phenomenon can be due to the passive aggregation of hemocytes around big structures or to the beginning of cellular encapsulation, which would then start earlier than what is detected by dissection.
Encapsulation was witnessed in one larva, sampled 48 hours after oviposition (Fig. 3, E, F). Inside the capsules, fully formed larvae are still protected by the egg chorion, and their morphology can be observed. The capsules appeared to be already several cells thick but were not finished, as the basal membrane could not be seen. Hemocytes were observed around the capsules, possibly recruited to form the layers.

### 3.5. Dynamic of hemocyte quantity

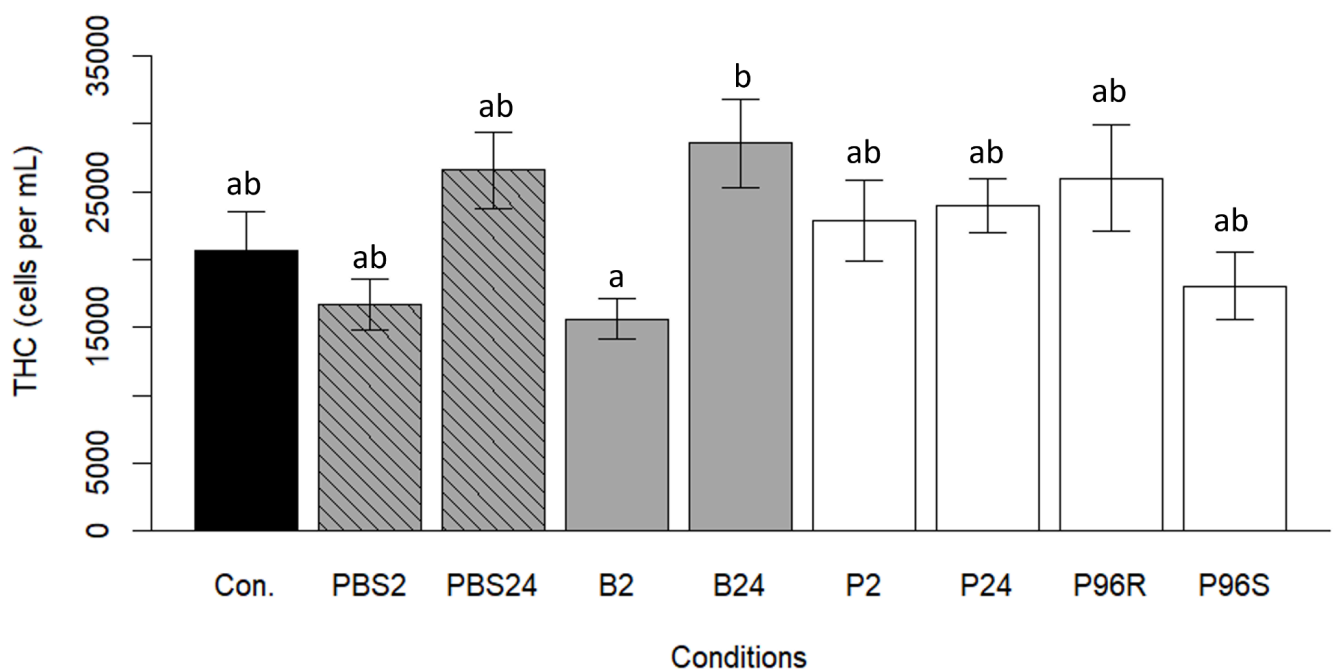

Fig. 4: Total hemocyte count of L5 *S. nonagrioides* for each condition. Con.: control larvae (black); PBS: sterile PBS (hatched gray); B: beads (gray); P: parasitization by Makindu strain (white). Numbers indicate the time step, expressed as hours post-treatment. For parasitization after 96 hours, results for resistant (R) or permissive (S) individuals are reported. Data represented are means of 8 individuals ± Standard Error of the Mean (SEM). They were compared by a one-way ANOVA followed by a post-hoc Tukey test. Different letters indicate a significant difference (p-adjusted < 0.05).

Hemocyte count was performed on 8 L5 SNF larvae for each condition to monitor the magnitude of the immune response to different challenges. The THC dynamic was measured in beads and sterile PBS-injected larvae at 2 and 24 hours, the time window of bead encapsulation. THC in parasitized larvae was measured at the same time points, but could not be categorized as resistant or permissive, as resistance is only visible from 48 hours after oviposition. THC in resistant and permissive larvae was thus assessed 96 hours (four days) after oviposition.

Total hemocyte count showed a significant effect of the treatments (Fig. 4, ANOVA: $F_{(8, 83)}$ = 2.898, p-value = 0.00816). Only the injection of beads had a significant effect, increasing the THC over time (B2 and B24, adjusted p-value = 0.0327). Injection of sterile PBS only did not influence THC, and no larvae died during the experiment, neither in the tested nor in the control groups, which excludes the effect of a mechanical wounding or bacterial infection on the observed THC. Parasitization also did not affect the THC, and there was no significant difference between resistant and permissive hosts after 96 hours post-oviposition.

## 4. Discussion

The main goal of this research was to characterize, through combined approaches, the encapsulation process of a parasitoid wasp, *Cotesia typhae* (Makindu strain), within a partially resistant host, *Sesamia nonagrioides* (French population).

### 4.1. Effect of C. typhae *parasitism on* S. nonagrioides *encapsulation process*

First, the dynamic of encapsulation was unraveled by successive dissections of resistant parasitized hosts at various times post-oviposition. This dynamic was compared to a positive control process by injecting inert chromatography beads in the caterpillars.

#### 4.1.1. Encapsulation dynamics

Results showed that the beads started to be encapsulated in less than an hour for every individual, and encapsulation was finished at 48 hours, with the presence of the basal membrane. This reaction can be rapid in other species (Götz & Boman, 1985; Havard et al., 2014; Uçkan et al., 2010), especially at the relatively high rearing temperature used in our experiments (Seehausen et al., 2017).

Encapsulation of the parasitoid started later, between 24 and 48 hours post-oviposition, a time during which the first cell layers gathered around the foreign body (Table 2). This encapsulation targeted the egg stage (Fig. 1, 2), consistently with what is reported in the literature in most host/parasitoid couples. No individual dissected before 48 hours presented encapsulated eggs or eggs with few hemocytes attached (Fig. 2). However, the presence of hemocytes around the eggs was observed at 24 and 36 hours post-oviposition through histological cuts (Fig. 3). Thus, histology can help refine the encapsulation dynamic, as the presence of few cells around the egg chorion would be challenging to assess by dissection. After four days post-oviposition, the capsule was complete in every resistant individual (Table 2), as could be seen with the presence of the basal membrane, which ends the encapsulation process (Fig. 1) and is possibly produced by granulocytes (Carton et al., 2008; Strand, 2008). This phenomenon prevents the formation of bigger clots that would harm the host larvae and can be used as a proxy for the speed of the immune reaction.

Basal membranes were observed after four days around encapsulated eggs against two days around beads (Table 2). Thus, the parasitoid encapsulation reaction is slower than that of the beads, even when beads are injected in parasitized larvae. In other species where resistance occurs more systematically, eggs appear to be encapsulated more quickly. The couple between the parasitoid *C. sesamiae* Mombasa (Hymenoptera: Braconidae) and the host *Busseola fusca* (Lepidoptera: Noctuidae) is characterized by complete host resistance (Ngi-Song et al., 1998). Gitau and collaborators (2007) showed that eggs started to be encapsulated 6 hours after oviposition, and the reaction was completed by 24 hours, which was very similar to encapsulation of the beads by *S. nonagrioides*. The same pattern was observed in three *Cotesia species* from the *flavipes* complex: *C. chilonis, C. sesamiae,* and *C. flavipes.* All eggs laid by each species were encapsulated in less than 24 hours by the completely resistant host *Ostrinia nubilalis* (Lepidoptera: Crambidae) (Alleyne & Wiedenmann, 2001). By contrast, French *S. nonagrioides* (SNF) resistance to *C. typhae* is known to be partial since it occurs in roughly 60 to 70 % of individuals (Benoist et al., 2020). Thus, this delayed reaction could result from a partial efficiency of the wasp virulence. In this hypothesis, immunity would be altered by parasitism and virulence factors, but successful encapsulation can still occur in more than half of the cases. The French *S. nonagrioides* population would be at a tipping point, where resistance can arise as likely as susceptibility and parasitoid virulence factors play a key role in the outcome of the interaction. *C. typhae* is a Microgastrinae wasp and thus injects polyDNAviruses alongside its eggs during the oviposition (Benoist et al., 2017; Herniou et al., 2013; Kaiser et al., 2017). Those viruses are known to alter the functioning of the host's immune system after infection of hemocytes (Beckage & Drezen, 2012; Strand & Burke, 2013) and to manipulate its physiology in order to facilitate parasitoid development (Gullan & Cranston, 2014). They could, therefore, not be efficient enough to alter sufficiently host immunity in all cases, and resistance could occur after a short delay, as it was hypothesized for the heavily delayed encapsulation of *C. sesamiae* and *C. flavipes* by *Diatraea grandiosella* (Lepidoptera: Crambidae) (Alleyne & Wiedenmann, 2001). Such a crucial role in between-strain virulence variation was demonstrated for *C. sesamiae* strains, where injection of virus from the virulent wasp in parasitized hosts restored the success of the avirulent strain (Mochiah et al., 2002). In endoparasitoid wasps, venom also plays an essential role in negating the host's immune system (Asgari, 2006), especially immediately after parasitism, during the time gap where viral particles have not integrated the host's cells yet and cannot produce virulence factors (Asgari & Rivers, 2011; Webb & Summers, 1990). In *Cotesia* species, venom is essential for viral expression in hemocytes (Moreau & Asgari, 2015). Thus, venom could be involved in delaying the encapsulation of Makindu eggs. Indeed, in *Galleria mellonella* (Lepidoptera: Pyralidae) and its parasitoid *Pimpla turionellae* (Hymenoptera: Ichneumonidae), injection of the wasp venom was sufficient to reduce encapsulation of beads by the caterpillar (Uçkan et al., 2010). Similarly, calyx fluid (containing polyDNAviruses) of *C. chilonis* delayed bead encapsulation in *Chilo suppressalis* (Lepidoptera: Crambidae) (Teng et al., 2016).

However, partial resistance can occur without encapsulation delay. *Hyposoter didymator* (Hymenoptera: Ichneumonidae) eggs were encapsulated within only 24 hours in resistant *Mythimna umbrigera* (Lepidoptera: Noctuidae), even if resistance occurred very scarcely (Bahena et al., 1999).

Another hypothesis explaining this consequent gap between bead and egg encapsulation would be that *C. typhae* eggs are protected from host immunity by passive evasion rather than by its impairment. In this kind of strategy, the host's immune system is left intact, but eggs cannot be detected as foreign bodies and, therefore, cannot trigger encapsulation. This hypothesis is consistent with the co-occurence of encapsulated beads and unencapsulated parasitoid eggs in SNF larvae parasitized by the Makindu strain. This result suggests that the immune system is unimpaired by the wasp's virulence factors. However, several non-encapsulated beads are still found in the hemolymph of some parasitized hosts and never in control hosts, indicating a partial action of immuno-disruptive factors injected by the wasp.

Passive evasion can be achieved by the presence of a sticky chorion around the eggs, which allows them to attach to the host's tissues and thus not be detected, as is the case for *Asobara tabida* (Kraaijeveld & Alphen, 1994; Prevost et al., 2005), or *Orgilus* (Hymenoptera: Braconidae) eggs (Quicke, 2014). However, histology experiments showed that *C. typhae* eggs were not embedded in particular host tissue, and most were observed to be freely dispersed in the hemolymph.

Several factors can protect the eggs from encapsulation, like glycoproteins brought to the surface of the eggs by virus-like particles (VLPs) in the wasp's oviduct (Feddersen et al., 1986). The resistant hosts can remove such surface factors at the expense of time, which would explain the delay in egg encapsulation (Huw Davies & Vinson, 1986). VLPs have not been reported in *C. typhae*, but surface factors could originate from the ovaries, as is the case for several *Cotesia* species (Teng et al., 2019). Passive evasion of *C. typhae* eggs remains to be explored but could contribute to both the encapsulation delay and the fact that resistance does not arise in every individual and on every egg.

### 4.1.2. Melanization levels

*C. typhae* egg capsules do not melanize (unless in host injected with beads), while bead capsules do (see Fig. 2), even if both are similarly constructed. Bead melanization appears to be weak, compared to melanization of wounds (Gornard, personnel observation). Weak melanization has already been observed for *Ephestia kuehniella* (Lepidoptera: Pyralidae) (Götz & Boman, 1985), *O. furnacalis* (Lepidoptera: Pyralidae), where it varies with age, or for *Amigeres subalbatus* (Diptera: Culicidae), where it depends on beads surface charge (Zahedi et al., 1992). Many factors injected alongside parasitism can negate this phenomenon, such as venom serine protease inhibitors (serpins), which negatively affect the melanization cascade by preventing the cleavage of key enzymes in *Drosophila* (Colinet et al., 2009; De Gregorio et al., 2002). Similarly, some viral proteins can target this cascade, such as Egf1.5, found in the bracovirus of *Microplitis demolitor* (Hymenoptera: Braconidae), a parasitoid of *Manduca sexta* (Lepidoptera: Bombycoidea) (Lu et al., 2010). While parasitism by *A. tabida* (Hymenoptera: Braconidae), a parasitoid of *Drosophila*, does not negatively affect its host immune system (Moreau et al., 2003), it still diminishes the phenoloxidase activity, the enzyme responsible for melanization. In this case, beads injected in parasitized larvae can be encapsulated (Doucet & Cusson, 1996; Hu et al., 2003) but do not melanize. Similarly, *C. chilonis* venom alone cannot alter encapsulation processes in *C. suppressalis* (Lepidoptera: Crambidae) but negatively affects melanization abilities (Teng et al., 2016). However, beads injected in parasitized larvae can still melanize. Therefore, parasitism by *C. typhae* does not seem to be able to alter the melanization abilities of its host, even in the close vicinity of the eggs, which precludes the presence of inhibitors attached to their surface. We observed that injection of beads induced melanization of parasitoid capsules. Since resistant larvae do not melanize parasitoid capsules when parasitism is the only immune challenge present, it is possible that detection of the eggs as non-self triggers only the encapsulation pathways, whereas challenging the host with beads would also trigger melanization cascades, and therefore all capsules would melanize.

### 4.2. Effect of parasitism on total hemocyte count

Capsule construction implies the recruitment of many hemocytes, and changes in the total hemocyte count (THC) of the host's hemolymph are expected in the case of bead injection and parasitism. THC increase is caused by recruitment from settled hemocytes or proliferation from free-roaming hemocytes or in dedicated hematopoietic organs. Successful parasitism can be expected to decrease THC by inhibiting this recruitment or even by causing apoptosis in order to negate the immune reaction (Beckage, 1998). Three days after parasitization of permissive *Choristoneura fumerana* (Lepidoptera: Torticidae) by *Tranosema rostrale* (Hymenoptera: Ichneumonidae), THC is half the one measured in non-parasitized larvae (Doucet & Cusson, 1996). Such a diminution is also observed in the permissive *Plutella xylostella* (Lepidoptera: Plutellidae) parasitized by *C. plutellae* (Hymenoptera: Braconidae) (Ibrahim & Kim, 2006). On the other hand, resistant hosts are expected to display an increased THC, reflecting the ongoing encapsulation processes toward the invader. The timing of the increase is variable. Resistant *Crocidolomia pavonana* (Lepidoptera: Crambidae) and *Spodoptera litura* (Lepidoptera: Noctuidae), parasitized by *Eriborus argenteopilosus* (Hymenoptera: Ichneumonidae) take up to 3 to 5 days to display increase in THC, even if encapsulation is completed within three days for *C. pavonana* (Buchori et al., 2009). At 24h, the authors did not observe significant changes in THC compared to the control. Alleyne & Wiedenmann (2001) reported a transient THC increase in resistant *Diatraea* species and *O. nubilalis* (respectively partially and totally resistant host) very shortly after parasitization by *C. flavipes*, which was back to control levels after 24 hours.

Changes in THC were compared when subjugating *S. nonagrioides* larvae to the same immune challenges as for establishing the encapsulation dynamic. Bead injection was expected to lead to an increase in THC in response to the immune reaction triggered by the invasion of foreign bodies.

Concerning bead-treated larvae, there was a significant increase between 2 and 24 hours post-injection (Fig. 4), likely linked to the encapsulation process that took place within this time gap, as expected. However, parasitized larvae showed a THC similar to that of the control, which did not change with either time or resistance status. No transient increase could be detected two and 24 hours after oviposition (Fig. 4). Ninety-six hours after oviposition, we could not detect any significant differences in THC, either between resistant and permissive larvae or compared to the control (although THC tended to be lower in permissive larvae, Fig. 4). At this time point, the encapsulation process is ongoing and almost finished in resistant larvae, and an increase in THC was expected in resistant larvae, alongside with a decrease in permissive ones. Thus, *C. typhae* parasitism does not seem to impact the THC of *S. nonagrioides*, consistent with what Alleyne & Wiedenmann (2001) reported for partially resistant *D. grandiosella* parasitized by *C. sesamiae* (sister species of *C. typhae*, Kaiser et al., 2017). In this case, as encapsulation of every parasitoid takes several days, the authors hypothesize that hemocyte recruitment in capsules is balanced by their production. This balance results in a stable equilibrium and could explain the absence of THC increase for resistant hosts at 96 hours. It is also consistent with the increased delay of parasitoid encapsulation when the host is already facing bead injection, compared to parasitism only (see Dynamic of encapsulation). This could be explained by the recruitment of hemocytes first towards bead encapsulation, and later towards egg encapsulation, upon recognition by the immune system. This shift could take time, involving recruitment or production of new hemocytes, and it explains the gap observed between egg encapsulation whether there are beads or not in the hemolymph.

On the other hand, the absence of THC decrease in permissive hosts can be explained by the passive evasion hypothesis. In this case, as there is no need to disrupt the immune system, THC remains constant and similar to the control during the first days of parasitism. Hu and collaborators (2003) showed that parasitism of *O. furnacalis* by *Macrocentrus cingulum* (Hymenoptera: Braconidae) did not alter the THC over time, only slightly decreasing it after five days post-parasitism. This could help understand the curious partial resistance of the French host to parasitism by the Makindu strain. It could be possible that only hosts with a sufficient hemocyte load can trigger the immune reaction by detecting the invasion of parasitoid eggs, like in *A. tabida* where the detection of concealed eggs depends on the number of hemocytes circulating in the drosophila larva (Eslin & Prevost, 1996; Prevost et al., 2005). If the detection of passively evading eggs depends on basal hemocyte count, this can imply an intermediate resistance level. In that case, an intermediate hemocyte load leads to the detection and encapsulation of only some eggs, explaining the lower offspring number sired by some yet permissive hosts, compared to parasitism on completely permissive host population (mean offspring number for Makindu strain: 62.53 on French host, 88.93 on Kenyan host, Gornard, personal communication). Partial encapsulation is indeed seen in one third of the resistant hosts (Table 1), and can lead to egression of fewer individuals.

Lack of THC change in permissive *S. nonagrioides* can also be linked to the effect of the venom in the Makindu strain. Endoparasitoid venom is known to play a critical role in hemocyte depletion and alteration (Asgari & Rivers, 2011; Mabiala-Moundoungou et al., 2010; Zhang et al., 2006) in species with impairing of the immune system. In our work, decrease in THC is not significant in permissive hosts, and the morphology of the hemocytes does not appear to be affected. Makindu strain venom cannot induce general apoptosis of the hemocytes, and this could be characteristic of *C. typhae* venom. Alternatively, it remains possible that *C. typhae* venom only provokes functional disruption without THC change, that is, preventing the cells from adhering, spreading, and recruiting others into the capsules. Other virulence factors can modulate this effect: *Chilo suppressalis* larvae exhibit higher THC when parasitized by *C. chilonis*, but the spreading and encapsulation ability was negated by the joint action of venom and calyx fluids (Teng et al., 2016). Some proteins synthesized by integrated polyDNAvirus also impede hemocyte spreading, such as the CrV1 protein of *C. rubecula* (Asgari & Schmidt, 2002). Thus, the effect of the different virulence factors must be thoroughly investigated to explain more precisely how parasitoid development can occur and why resistance arises in some cases.

## 5. Conclusion

By combining several approaches, we could characterize both parasitoid development and encapsulation dynamics inside a partially resistant host. Those results provide us with a precise time frame for collecting tissues to perform molecular analysis on the virulence factors of the wasps and the resistance factors of the host. Comparing the bead encapsulation process to that of parasitoid eggs, we conclude that the Makindu strain of *C. typhae* resorts to a passive evasion strategy when parasitizing its host. Here, the host's immune system remains functional and able to encapsulate and melanize inert beads.This is coherent with the low variation of hemocyte numbers following parasitism compared to the injection of beads. *Cotesia typhae* and *Sesamia nonagrioides*, species of agro-economic interests displaying both permissive and forbidding interactions, provide a very interesting model to understand more precisely the interactions between virulence strategies and resistance mechanisms.

## Funding

This study was co-funded by the French National Research Agency (ANR) and the National Agency for Biodiversity (AFB) (grant CoteBio ANR17-CE32-0015-02 to L.K.) and by the Ecole doctorale 227 MNHN-UPMC Sciences de la Nature et de l'Homme: évolution et écologie

## CRediT authorship contribution statement

**Samuel Gornard**: Conceptualization, Data curation, Formal analysis, Investigation, Writing - original draft. **Florence Mougel**: Conceptualization, Formal analysis, Project administration, Resources, Supervision, Validation, Writing - review & editing. **Isabelle Germon**: Conceptualization, Data curation, Investigation, Methodology, Writing - review & editing. **Véronique Borday-Birraux**: Conceptualization, Data curation, Investigation, Methodology, Resources, Writing - review & editing. **Pascaline Venon**: Investigation. **Salimata Drabo**: Investigation. **Laure Kaiser**: Conceptualization, Funding acquisition, Project administration, Resources, Supervision, Validation, Visualisation, Writing - review & editing

## Declaration of Competing Interest

The authors declare that they have no known competing financial interests or personal relationships that could have appeared to influence the work reported in this paper.

# Data availability

# Acknowledgments

We thank Alain Peyhorgues for providing field-collected *S. nonagrioides* individuals for laboratory rearing at EGCE; Rémi Jeannette for *S. nonagrioides* rearing; Claire Capdevielle-Dulac and Sabrina Bothorel for *C. typhae* rearing at EGCE, Paul-Andre Calatayud and the Cotesia rearing team at ICIPE (Nairobi, Kenya) - Julius Obonyo, Josphat Akhobe, and Enock Mwangangi - for insect collection and shipment to refresh EGCE rearing; Pauline Depierrefixe for her help in laboratory work.